# An Unsupervised Approach for Discovering Relevant Tutorial Fragments for APIs

He Jiang[1, 2, 3]
jianghe@dlut.edu.cn

Jingxuan Zhang[1]
jingxuanzhang@mail.dlut.edu.cn

Zhilei Ren[1]
zren@dlut.edu.cn

Tao Zhang[4]
cstzhang@hrbeu.edu.cn

[1]School of Software, Dalian University of Technology, Dalian, China
[2]Key Laboratory for Ubiquitous Network and Service Software of Liaoning Province, Dalian, China
[3]State Key Laboratory of Software Engineering, Wuhan University, Wuhan, China
[4]College of Computer Science and Technology, Harbin Engineering University, Harbin, China

*Abstract*—Developers increasingly rely on API tutorials to facilitate software development. However, it remains a challenging task for them to discover relevant API tutorial fragments explaining unfamiliar APIs. Existing supervised approaches suffer from the heavy burden of manually preparing corpus-specific annotated data and features. In this study, we propose a novel unsupervised approach, namely Fragment Recommender for APIs with PageRank and Topic model (FRAPT). FRAPT can well address two main challenges lying in the task and effectively determine relevant tutorial fragments for APIs. In FRAPT, a Fragment Parser is proposed to identify APIs in tutorial fragments and replace ambiguous pronouns and variables with related ontologies and API names, so as to address the *pronoun and variable resolution* challenge. Then, a Fragment Filter employs a set of non-explanatory detection rules to remove non-explanatory fragments, thus address the *non-explanatory fragment identification* challenge. Finally, two correlation scores are achieved and aggregated to determine relevant fragments for APIs, by applying both topic model and PageRank algorithm to the retained fragments. Extensive experiments over two publicly open tutorial corpora show that, FRAPT improves the state-of-the-art approach by 8.77% and 12.32% respectively in terms of *F-Measure*. The effectiveness of key components of FRAPT is also validated.

## I. INTRODUCTION

Reusing Application Programming Interfaces (APIs) in existing libraries and frameworks could greatly speed up software development process [1, 2]. Hence, it is critical for developers to search and use APIs properly [3]. As one of the most important API documentations, API tutorials are often consulted by developers to learn how to use APIs in a given programming context [4, 5]. However, API tutorials are usually lengthy and mixed with irrelevant information, so it is tedious for developers to peruse a full tutorial for an unfamiliar API [6]. One feasible solution is to split tutorials into consecutive fragments and recommend relevant fragments containing explanation information to developers [3, 7]. Hereafter, if a fragment explains an API, then they are relevant. Otherwise, they are irrelevant.

In recent years, three approaches have been proposed to resolve the task of discovering relevant tutorial fragments for APIs in the literature, namely FITSEA [7], GMR [3], and IR [3]. Among these approaches, FITSEA and GMR are supervised approaches and dramatically outperform IR. There are two stages in these supervised approaches, namely training and test. In the training stage, tutorials are split into fragments and APIs are manually labeled as relevant or irrelevant to their fragments. The labeled APIs and their fragments form a series of fragment-API pairs. Some predefined features are extracted from each fragment-API pair. These extracted features with their class labels (relevant or irrelevant) are taken to train a classifier. In the test stage, the trained classifier is used to predict the class labels of new fragment-API pairs.

However, there are some drawbacks in these supervised approaches, due to the heavy burden of manual annotation efforts and feature construction.

● Different corpora require their corpus-specific annotated data, thus cost a lot of manual efforts. Supervised approaches may easily lead to the bias between training and test, especially when conducting cross-corpus prediction [3].

● The effectiveness of supervised approaches largely depends on the features to capture the specific characteristics of corpora. When applying the constructed features on different corpora, classifiers may behave well on some corpora, while poorly on the others [3, 7].

The above drawbacks in supervised approaches motivate us to propose an unsupervised approach, which possesses several advantages compared to supervised approaches [8, 9]. First, there is no need to annotate corpus-specific data and train a classifier, so that it costs far less manual efforts and avoids the bias between training and test. Second, unsupervised approaches can adapt to different corpora by adjusting some parameters instead of constructing features.

In this study, we propose a novel unsupervised approach, namely Fragment Recommender for APIs with PageRank and Topic model (FRAPT). FRAPT can find the relevance between fragments and APIs effectively and recommend relevant fragments for APIs to developers automatically. In FRAPT, all fragments are taken into a Fragment Parser to identify APIs, replace ambiguous pronouns and variables with related ontologies and API names, and find sentence boundaries and their types after tutorial segmentation. Then, a Fragment Filter is introduced to filter out non-explanatory fragments which do not explain any APIs. Next, both topic model and PageRank algorithm are applied to each retained fragment to achieve two types of correlation scores between fragments and APIs. By aggregating the correlation scores, relevant fragments for APIs can be identified and recommended for developers.

To evaluate the effectiveness of FRAPT and validate the impacts of its key components, we conduct extensive experiments on all publicly available annotated tutorial corpora, namely McGill corpus [38] and Android corpus [39]. FRAPT improves the state-of-the-art approach FITSEA by 8.77% and 12.32% respectively in terms of *F-Measure*. In addition, the key components of FRAPT also exhibit their effectiveness. For example, by aggregating the correlation scores from both topic model and PageRank algorithm, better *F-Measure* can be achieved and improve any of them by up to 9.10% and 4.88% over the two tutorial corpora.

The main contributions of this study are listed as follows:

(1) A Fragment Filter with a set of non-explanatory detection rules is proposed to remove non-explanatory fragments effectively.

(2) We propose a new unsupervised approach FRAPT leveraging topic model and PageRank algorithm to discover relevant fragments for APIs. FRAPT is publicly available at http://oscar-lab.org/people/~jxzhang/FRAPT/.

(3) A series of experiments are conducted to validate the effectiveness of FRAPT and its components. The results demonstrate that FRAPT significantly improves the state-of-the-art supervised approach.

## II. MOTIVATION EXAMPLES

In this section, we discuss the main challenges lying in the task of discovering relevant fragments for APIs.

APIs can be explained at different levels by tutorials, ranging from package level, class or interface level, to method level. In order to facilitate comparison of different approaches, similar as [3, 7], APIs are chosen at class or interface level in this study. On one hand, when facing an unfamiliar API, developers usually want to know how to use some available behaviors rather only one method [3]. On the other hand, tutorials usually introduce some programming topics by using a set of methods of classes [7]. As a result, APIs are selected at class or interface level in this study.

Fig. 1 shows two fragment examples. Fragment *A* and Fragment *B* are extracted from the JodaTime tutorial [40] and the Android Graphics tutorial [41] respectively. Fragment *A* has three paragraphs with four APIs, namely *DateTime*, *Interval*, *Duration*, and *Period*. However, Fragment *A* is irrelevant to any of them according to the manual annotation [40]. In contrast, there are two paragraphs with a piece of code example in Fragment *B*. Four APIs are detected, namely *Canvas*, *Bitmap*, *SurfaceHolder*, and *SurfaceView*. According to the manual annotation [41], *Canvas* and *Bitmap* are relevant to this fragment, whereas *SurfaceHolder* and *SurfaceView* are not.

Based on the above tutorial fragments, we find that two challenges need to be addressed for effectively discovering relevant tutorial fragments for APIs.

● *Pronoun and variable resolution*

Pronouns and variables are widely used in API tutorials. These pronouns and variables may be ambiguous, if we do

```
(1)JodaTime is like an iceberg, 9/10ths of it is invisible to user-code. (2)Many, perhaps most, applications will never need to see what's below the surface.
(3)This document provides an introduction to the JodaTime API for the average user, not for the would-be API developer.
(4)The bulk of the text is devoted to code snippets that display the most common usage scenarios in which the library classes are used. (5)In particular, we cover the usage of the key DateTime, Interval, Duration and Period classes.
(6)We finish with a look at the important topic of formatting and parsing and a few more advanced topics.
```
(a) Fragment *A*

```
(1)When you're writing an application in which you would like to perform specialized drawing and/or control the animation of graphics, you should do so by drawing through a Canvas. (2)A Canvas works for you as a pretense, or interface, to the actual surface upon which your graphics will be drawn. (3)It holds all of your "draw" calls. (4)Via the Canvas, your drawing is actually performed upon an underlying Bitmap, which is placed into the window.
(5)In the event that you're drawing within the onDraw() callback method, the Canvas is provided for you and you need only place your drawing calls upon it.
(6)You can also acquire a Canvas from SurfaceHolder.lockCanvas(), when dealing with a SurfaceView object (Both of these scenarios are discussed in the following sections). (7)However, if you need to create a new Canvas, then you must define the Bitmap upon which drawing will actually be performed. (8)The Bitmap is always required for a Canvas. (9)You can set up a new Canvas like this:

(10) Bitmap b = Bitmap.createBitmap(100, 100, Bitmap.Config.ARGB_8888);
(11) Canvas c = new Canvas(b);
```
(b) Fragment *B*

Figure 1. Fragment examples

not consider their surrounding sentences or programming context. We observe that about 70% fragments contain at least one pronoun and 50% fragments with code examples declare variables in the open tutorial corpora [38, 39]. The influences of APIs will be weakened, if the ontologies of these pronouns and variables are APIs.

Taking Fragment *B* from the Graphics tutorial [41] in Fig. 1 as an example, if we only consider the third sentence "*It holds all of your draw calls*", the ontology of "*It*" is ambiguous. With the help of its context, we can infer that "*It*" stands for the API *Canvas*. In contrast, the second statement in code example in Fragment *B* declares a *Canvas* variable *c* with *b* as a parameter. By inspecting its programming context, we can infer that *b* stands for the API *Bitmap*.

● *Non-explanatory fragment identification*

Not all the fragments are designed to explain APIs. Non-explanatory fragments do not explain any APIs (accordingly, an explanatory fragment explains at least one API). As to an investigation on the tutorial corpora used in this study [38, 39], we find that 10% ~ 50% fragments are annotated as non-explanatory fragments.

For example, Fragment *A* belongs to non-explanatory fragments, since all of its APIs appear in an enumeration sentence, and Fragment *A* just gives an overview of the whole JodaTime tutorial [40].

In the following part of this paper, we present how our new approach FRAPT addresses the above challenges and explain its workflow with Fragment A and Fragment B.

## III. FRAMEWORK OF FRAPT

The whole framework of FRAPT consists of two phases, namely the relevance discovery phase and the fragment recommendation phase (see Fig. 2).

**Relevance Discovery Phase.** This phase aims to find the relevance between fragments and APIs. FRAPT first segments tutorials into coherent fragments by a Tutorial Segmentation component, and inputs the fragments into a Fragment Parser to detect APIs, resolve pronouns in sentences and variables in code examples to address the *pronoun and variable resolution* challenge, and identify sentences with their types. Then, the parsed fragments are taken into a Fragment Filter to detect and filter out non-explanatory fragments, so as to address the *non-explanatory fragment identification* challenge. Next, on one hand, the fragments are taken into a topic model to find semantic connections between fragments and APIs. Based on the fragment-topic matrix and the topic-term matrix obtained from the topic model, a type of correlation scores between fragments and APIs can be achieved. On the other hand, PageRank algorithm is applied to the graph constructed by sentences in fragments, and another type of correlation scores between fragments and APIs can also be achieved according to the PageRank values of sentences containing APIs. Last, a relevance identification scheme is designed to decide the relevance between fragments and APIs.

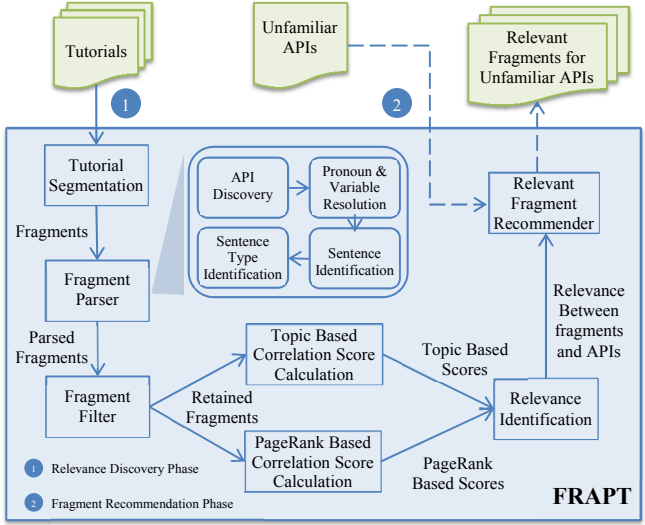

Figure 2. Workflow of FRAPT

**Fragment Recommendation Phase.** The goal of this phase is to recommend relevant fragments for the unfamiliar APIs queried by developers. When a developer wants to use an unfamiliar API to accomplish some programming tasks, he/she inputs the unfamiliar API into FRAPT. FRAPT looks up the relevance between fragments and APIs obtained from the first phase, and recommends relevant tutorial fragments to the developers.

## IV. MAIN COMPONENTS OF FRAPT

### A. Tutorial Segmentation

Tutorial segmentation follows the same method as [3, 7] to divide tutorials into coherent fragments. First, we split tutorials into a series of paragraphs. Then, we iteratively merge sibling paragraphs until the length of the paragraphs reaches within a specified range (100 words to 300 words). In such a way, the tutorial fragments can be achieved for each tutorial.

### B. Fragment Parser

Fragment Parser conducts four steps to parse a fragment, namely API Discovery, Pronoun and Variable Resolution, Sentence Identification, and Sentence Type Identification.

*(1) API Discovery*

First of all, we detect all the APIs contained in each fragment. All the open tutorials are available on the Internet in the form of HTML. Although APIs in different tutorials have different styles, most of tutorials follow W3C guidelines and link APIs to their corresponding specification webpages with <href> tags. By parsing the linked webpages and matching API names with the anchor text, we can decide whether a link is an API or not.

Some tutorials do not link APIs to their specification webpages. In this situation, we tokenize the tutorials into a stream of tokens and match these tokens with all the available API names. The same as [19], if a token matches an API name lexically, we continue to search its context, a sequence of tokens surround this token with a window size of 2, to find whether it contains some keywords, such as "*class*", "*interface*", and "*API*". If yes, it is treated as an API.

With API discovery, four APIs are detected for each fragment in Fig. 1. *DateTime*, *Interval*, *Duration*, and *Period* are discovered in Fragment *A*. *Canvas*, *Bitmap*, *SurfaceHolder*, and *SurfaceView* are detected in Fragment *B*.

*(2) Pronoun and Variable Resolution*

To address the *pronoun and variable resolution* challenge, Pronoun and Variable Resolution identifies ontologies for pronouns and APIs for variables, and replaces pronouns and variables with their ontologies and APIs.

In FRAPT, Reconcile [10], an automatic coreference resolution tool, is leveraged to perform pronoun resolution, thus we can obtain mappings between pronouns and their ontologies. We replace all the pronouns with their ontologies for all the sentences. In contrast, if a variable is declared in a piece of code example and some operations are performed on the variable in the following statements, the variable is used to call its methods rather than its API, especially in object-oriented programming languages. We inspect each statement in the code examples, and build a mapping between each variable and its API. Then we replace all the variables with their APIs for each statement. In such a way, the impacts of APIs are enhanced.

Taking Fragment *B* as an example, the third sentence "*It holds all of your 'draw' calls.*" in the first paragraph starts with a pronoun, namely "*It*". By analyzing the context of this sentence, Reconcile [10] finds that "*It*" stands for the API *Canvas*. As a result, we replace "*It*" with *Canvas* so that the API *Canvas* is enhanced. There are two code statements in this fragment. The first statement creates a *Bitmap* variable *b*, and the second statement declares a *Canvas* variable *c* using *b* as a parameter. By analyzing the statements, we can create two mappings, namely <*b*, *Bitmap*> and <*c*, *Canvas*>. Variable *b* appears in the second statement, and we replace it with its API, namely *Bitmap*. As a result, the two statements are changed to:

```
(10) Bitmap b = Bitmap.createBitmap(100, 100, Bitmap.Config.ARGB_8888);
(11) Canvas c = new Canvas(Bitmap);
```

*(3) Sentence Identification*

Tutorials usually contain not only API explanations, but also code examples to show how APIs are used in specific situations. To obtain each sentence, we delete all the HTML tags in fragments. An open tool, namely LingPipe [11] is leveraged to detect the sentence boundaries so that the sentences are identified. We treat each statement in code examples as one sentence. We can obtain statements by splitting at the places of semicolons. However, there are some exceptions, *e.g.*, "*if*" and "*for*" statements. For the exceptions, we can split code at this place where there is a match between left parenthesis and right parenthesis.

We identify 6 and 11 sentences for Fragment *A* and Fragment *B* respectively, and mark each sentence with a sentence ID (see Fig. 1).

*(4) Sentence Type Identification*

Different types of sentences have different contributions to explain APIs. To differentiate the importance of each sentence, we define two sentence types, *i.e.* marginal sentences and principal sentences. Marginal sentences include conditional sentences, enumeration sentences, example sentences, comparative sentences, and code comments. Principal sentences are the other sentences except the above sentences. Intuitively, principal sentences are more important than marginal sentences. Marginal sentences can be detected by keyword matching in the following ways:

● Conditional sentences can be found by keywords of "*if*" and "*whether*".

● Enumeration sentences can be detected, when more than two APIs are listed and connected by "*and*" or "*or*".

● Example sentences can be detected by some keywords, *e.g.*, "*for example*", "*for instance*", and "*such as*".

● Comparative sentences can be detected by keywords such as "*compared to*", "*unlike…*", and "*more…than…*". In this study, inheritance-describing sentences are also treated as comparative sentences, which can be found by keywords, *e.g.*, "*extend*", "*inherited from*", and "*subtype of*".

● Code comments start with the string of "*//*", "*/\**" etc.

The fifth sentence in Fragment *A* is an enumeration sentence and the seventh sentence in Fragment *B* is a conditional sentence. As a result, they are marginal sentences. The other sentences can be viewed as principal sentences.

*C. Fragment Filter*

Fragment Filter aims to filter out those non-explanatory fragments which do not explain any APIs by some non-explanatory detection rules. We first specify how we perform an in-depth observation to identify these non-explanatory detection rules. Then we show the details of the rules.

*(1) Non-explanatory Detection Rules Identification*

There are three steps to identify non-explanatory detection rules, namely non-explanatory fragment identification, characteristics formulation, and non-explanatory detection rules generation.

**Non-explanatory Fragment Identification.** We select the Essential Java Classes tutorials as the representative tutorials to observe, since they introduce basic Java classes (*e.g.*, exception handing and I/O processing), and do not overlap with the two tutorial corpora used in this study [37]. Therefore, the identified non-explanatory detection rules can be applied to the tutorial corpora in this study without introducing a sampling bias into the results. We segment the Essential Java Classes tutorials into fragments by the Tutorial Segmentation component, and evenly distribute these fragments to the second and the third author of this study to check. They are required to analyze and understand these fragments to select the non-explanatory fragments.

**Characteristics Formulation.** For each identified non-explanatory fragment, the two authors investigate what are the characteristics that lead this fragment to be a non-explanatory fragment. To find the answer, the authors measure different characteristics for each non-explanatory fragment, and check if one or some of them could answer this question, *e.g.*, the number of sentences and the number of contained APIs. If they convince that one or some characteristics of a fragment make it to be a non-explanatory fragment, they formulate a new characteristic or merge it to an existing one. Finally, the authors achieve two sets of formulated characteristics.

**Non-explanatory Detection Rules Generation.** The two authors check the two formulated characteristic sets to find the common characteristics appearing in both two sets. Inter-rater Kappa agreement is applied to evaluate the two formulated characteristic sets [36]. The Kappa agreement is 0.46 showing a moderate agreement. Then, they decide the most suitable threshold for each common characteristic found by two authors simultaneously. For example, how many sentences can differentiate non-explanatory fragments from explanatory fragments. In such a way, through successive iterations of discussion and coordination, they find some characteristics that could well detect the boundary between non-explanatory fragments and explanatory fragments. These characteristics with their thresholds are transformed to some non-explanatory detection rules.

*(2) Non-explanatory Detection Rules*

After the non-explanatory detection rules identification, we find that non-explanatory fragments comply with at least one of the following characteristics (rules):

● **Non-explanatory fragments usually contain only one API, and this API appears only once.**

The more times an API appears in a fragment, the more chances the fragment is relevant to this API. The fragment pays less attention on API explanation, if there is only one API.

● **The lengths of non-explanatory fragments are usually less than five sentences.**

In such a short fragment, an API cannot be fully explained, considering that most of APIs may contain more than one method so as to have sophisticated behaviors.

● **The proportion of sentences containing APIs is less than 20% in non-explanatory fragments.**

It is hard to explain any API in such a small proportion of sentences containing APIs.

● **All the APIs in non-explanatory fragments only appear in marginal sentences.**

In this situation, APIs are only used to be listed as examples or enumeration. In the other principal sentences, the fragments focus on other information rather than APIs.

Table I. API PROBABILITIES ASSIGNED TO EACH TOPIC

| Term (API) | Topic 1 | Topic 2 | Topic 3 | Topic 4 | Topic 5 | … |
|---|---|---|---|---|---|---|
| canvas | 0.0002 | 0.0360 | 0.0000 | 0.4270 | 0.0000 | … |
| bitmap | 0.0000 | 0.0000 | 0.0000 | 0.3360 | 0.0005 | … |
| surfaceholder | 0.0001 | 0.0481 | 0.0000 | 0.0054 | 0.0000 | … |
| surfaceview | 0.0023 | 0.0336 | 0.0000 | 0.0005 | 0.0000 | … |

- **Non-explanatory fragments have some good indicator terms and phrases.**

Some keywords and phrases can give important clues and act as good indicators for finding non-explanatory fragments, e.g., "*summary*", "*overview*", "*introduction*", "*for more information about ..., see ...*", and "*more details in ...*".

Based on these characteristics, we construct a Fragment Filter with five non-explanatory detection rules to identify non-explanatory fragments. If one fragment conforms to any of these rules, it is marked as a non-explanatory fragment to be filtered out.

In Fragment *A*, all the APIs appear in an enumeration sentence, so it follows the fourth rule. As a result, Fragment *A* is detected as a non-explanatory fragment and filtered out. In contrast, Fragment *B* is inconsistent with none of these rules and requires further processing.

### D. Topic Based Correlation Score Calculation

Topic based correlation score calculation aims to achieve a correlation score between fragments and APIs according to the results of topic model. As a popular way to analyze a large scale of documents, topic model can be used to find semantic relationships between documents and terms [12, 13]. As a result, each document is expressed by a series of topics with different probabilities. In contrast, a topic is represented by a collection of terms with various probabilities. In this study, one type of topic model, namely Latent Dirichlet Allocation (LDA) is leveraged, and Stanford Topic Modeling Toolbox [14] is introduced to help us to perform LDA. We introduce the same method as [12] to obtain the best configurations (i.e., the topic number) for LDA. Each tutorial fragment is treated as a document to put into LDA, and LDA outputs the topic probability distributions for each fragment as well as the term probability distributions for each topic. Here, the terms include not only pure text, but also APIs. Since we intend to investigate the correlation scores between fragments and APIs, only the API probability distributions for all the topics are used in the following part.

Based on the output of LDA, we can obtain a correlation score between fragments and APIs by the following formula:

$$\text{Score}_T(\text{API, fragment}) = \sum_{t \in \text{topic}} P(\text{API} \mid t) \times P(t \mid \text{fragment}) \quad (1)$$

where $P(\text{API} \mid t)$ is the API probability for topic t, and $P(t \mid \text{fragment})$ is the probability of topic t for the fragment.

Taking Fragment *B* as an example, the fragment can be represented by five topics with non-zero probabilities, namely {0.04, 0.01, 0.02, 0.92, 0.01}. Table I shows the API probability distributions on the five topics. Following the formula (1), the $\text{Score}_T$ values for *Canvas*, *Bitmap*, *SurfaceHolder*, and *SurfaceView* with Fragment *B* are 0.39321, 0.30913, 0.00545, and 0.00089 respectively.

TABLE II. SIMILARITY MATRIX OF DIRECTED GRAPH FOR FRAGMENT *B*

| ID | (1) | (2) | (3) | (4) | (5) | (6) | (7) | (8) | (9) | (10) | (11) |
|---|---|---|---|---|---|---|---|---|---|---|---|
| (1) | - | 0.1573 | 0.1759 | 0.2580 | 0.2018 | 0.0706 | 0.3419 | 0.1207 | 0.2060 | 0.0000 | 0.2006 |
| (2) | 0.1573 | - | 0.1015 | 0.0766 | 0.0782 | 0.1465 | 0.1015 | 0.1252 | 0.2137 | 0.0000 | 0.2081 |
| (3) | 0.1759 | 0.1015 | - | 0.1628 | 0.3227 | 0.1058 | 0.2157 | 0.1808 | 0.3086 | 0.0000 | 0.3005 |
| (4) | 0.2580 | 0.0766 | 0.1628 | - | 0.1697 | 0.0799 | 0.5480 | 0.3802 | 0.2330 | 0.2423 | 0.4660 |
| (5) | 0.2018 | 0.0782 | 0.3227 | 0.1697 | - | 0.0815 | 0.2249 | 0.1394 | 0.2379 | 0.0000 | 0.2317 |
| (6) | 0.0706 | 0.1465 | 0.1058 | 0.0799 | 0.0815 | - | 0.1058 | 0.1306 | 0.2227 | 0.0000 | 0.2169 |
| (7) | 0.3419 | 0.1015 | 0.2157 | 0.5480 | 0.2249 | 0.1058 | - | 0.5038 | 0.3087 | 0.3864 | 0.6175 |
| (8) | 0.1207 | 0.1252 | 0.1808 | 0.3802 | 0.1394 | 0.1306 | 0.5038 | - | 0.3809 | 0.3961 | 0.7618 |
| (9) | 0.2060 | 0.2137 | 0.3086 | 0.2330 | 0.2379 | 0.2227 | 0.3087 | 0.3809 | - | 0.0000 | 0.6329 |
| (10) | 0.0000 | 0.0000 | 0.0000 | 0.2423 | 0.0000 | 0.0000 | 0.3864 | 0.3961 | 0.0000 | - | 0.3887 |
| (11) | 0.2006 | 0.2081 | 0.3005 | 0.4660 | 0.2317 | 0.2169 | 0.6175 | 0.7618 | 0.6329 | 0.3887 | - |
| PR | **0.8018** | **0.6062** | **0.8563** | **1.1264** | **0.7846** | **0.5857** | **1.4106** | **1.3163** | **1.1869** | **0.6559** | **1.6694** |

### E. PageRank Based Correlation Score Calculation

This component aims to obtain another correlation score between fragments and APIs based on the PageRank value of each sentence. PageRank algorithm attempts to evaluate the importance of each sentence through link analysis [15, 16]. First of all, we create a directed graph based on the sentences in a fragment, and each vertice in the graph stands for a sentence. To build directed edges between vertices, we calculate the widely used cosine similarity with TF-IDF term weight between sentences [7]. If the similarity is not zero between two sentences, we draw two directed edges toward each other between their corresponding vertices. In such a way, we can build a directed graph. After PageRank algorithm is applied to the directed graph, we can obtain the PageRank value for each sentence.

Different sentences have different contributions to explain APIs. The greater the PageRank value of a sentence, the more closely it catches the central meaning of the fragment. If an API appears in sentences with greater PageRank values, then it is more likely to be explained by this fragment. A scheme is designed to obtain correlation scores between fragments and APIs as follows:

$$\text{Score}_{PR}(\text{API, fragment}) = \frac{\sum_{s \in \text{fragment}} PR(s) \times P(\text{API} \mid s)}{\sum_{s \in \text{fragment}} PR(s)} \quad (2)$$

where s stands for a sentence in the fragment, PR (s) means the PageRank value of sentence s, and

$$P(\text{API} \mid s) = \begin{cases} 1 & \text{if s contains API} \\ 0 & \text{if s does not contain API} \end{cases} \quad (3)$$

Table II shows the matrix representation of the directed graph with PageRank values listed in the last line for Fragment *B*. The IDs are their corresponding sentence ID, and the values in the table show the similarities between the corresponding sentences. The PageRank values range from 0.5857 to 1.6694 for each sentence. The API *Canvas* appears in ten sentences. Note that *Canvas* has already replaced "*It*" in the sentence "*It holds all of your 'draw' calls.*" after pronoun resolution. The $\text{Score}_{PR}$ between *Canvas* and Fragment *B* is 0.9404 according to the formula (2). *Bitmap* appears five times, since the variable *b* in the last statement has been replaced by *Bitmap*. As a result, the $\text{Score}_{PR}$ between *Bitmap* and Fragment *B* is 0.5617. In the same way, the $\text{Score}_{PR}$ values for *SurfaceHolder* and *SurfaceView* with Fragment *B* are all 0.0532.

Table III. RELEVANCE IDENTIFICATION SCHEME

| | |
|---|---|
| **Input:** sentence types, $Score_T$ and $Score_{PR}$ values between APIs and their Fragment F, threshold T | |
| **Output:** the relevance between fragments and APIs; | |
| 1 | **for each** (API A in F) |
| 2 | normalize $Score_T(A, F)$ and $Score_{PR}(A, F)$; |
| 3 | find all sentences containing A; |
| 4 | **if** (types of these sentences are all marginal sentences) |
| 5 | A is irrelevant to F; |
| 6 | **else** |
| 7 | **if** (normalized $Score_T(A, F)$>=T&&normalized $Score_{PR}(A, F)$>=T) |
| 8 | A is relevant to F; |
| 9 | **else** |
| 10 | A is irrelevant to F; |

Table IV. DETAILS OF THE TUTORIAL CORPORA

| Corpus | Tutorial | API | Explanatory fragment | Non-explanatory fragment | Fragment with code | Fragment without code |
|---|---|---|---|---|---|---|
| McGill Corpus [38] | JodaTime | 36 | 19 | 10 | 21 | 8 |
| | Math Library | 73 | 31 | 10 | 16 | 25 |
| | Col. Official | 59 | 31 | 26 | 17 | 40 |
| | Col. Jenkov | 28 | 34 | 35 | 42 | 27 |
| | Smack | 40 | 42 | 5 | 31 | 16 |
| Android Corpus [39] | Graphics | 70 | 26 | 12 | 19 | 19 |
| | Resources | 63 | 40 | 6 | 34 | 12 |
| | Data | 37 | 18 | 7 | 9 | 16 |
| | Text | 31 | 12 | 12 | 10 | 14 |

*F. Relevance Identification*

We design a relevance identification scheme to detect the relevance between fragments and APIs.

In this study, we introduce a threshold *T* for each tutorial as one of the inputs of relevance identification scheme, which is used by all the fragments in the tutorial, to compare with the two correlation scores to decide the relevance between fragments and APIs. In the experimental results section (see VI.A), we will demonstrate how the threshold *T* influences the results and how to find a good value of it automatically.

Table III shows the relevance identification scheme. For each API in a fragment, we normalize the two correlation scores between this API and its fragment, namely each score is divided by the maximum score of the same type of scores in this fragment respectively. Then, we check whether all the types of sentences containing this API are marginal sentences. If yes, then we treat that this API only appear in less important sentences, and mark it as irrelevant to this fragment. If the two normalized correlation scores between this API and its fragment, namely $Score_T$ and $Score_{PR}$, exceed the threshold *T* at the same time, then the API is identified as relevant to the fragment. Otherwise, the API is identified as irrelevant to this fragment. With such a scheme, we can find the relevance between fragments and APIs.

Still taking Fragment *B* as an example, $Score_T$ and $Score_{PR}$ values for *Canvas*, *Bitmap*, *SurfaceHolder*, and *SurfaceView* with Fragment *B* are {1, 0.7862, 0.0139, 0.0023} and {1, 0.5973, 0.0566, 0.0566} respectively after normalization. None of these APIs only appear in marginal sentences, so they need further processing. If the threshold *T* is set to 0.29, we can find that both $Score_T$ and $Score_{PR}$ of *Canvas* and *Bitmap* with Fragment *B* are greater than the threshold *T* simultaneously. Hence, they are marked as relevant to Fragment *B*. Neither $Score_T$ nor $Score_{PR}$ of *SurfaceHolder* and *SurfaceView* with Fragment *B* exceed the threshold *T*, so they are marked as irrelevant to Fragment *B*.

In summary, with FRAPT, Fragment *A* is marked as a non-explanatory fragment and filtered out. *Canvas* and *Bitmap* are relevant to Fragment *B*, whereas *SurfaceHolder* and *SurfaceView* are not.

*G. Relevant Fragment Recommender*

In the fragment recommendation phase, Relevant Fragment Recommender recommends relevant fragments for unfamiliar APIs queried by developers by looking up the detected relevance between fragments and APIs. For example, if a developer inputs an unfamiliar API *Canvas* to FRAPT. The recommender finds that Fragment *B* is relevant to *Canvas* and recommends Fragment *B* to the developer.

V. EXPERIMENTAL SETUP

*A. Tutorial Corpora*

In the previous studies [3, 7], two publicly open tutorials are constructed shown in Table IV, which shows the number of APIs and fragments rather than fragment-API pairs. McGill corpus consists of five tutorials and Android corpus is composed of four tutorials. These corpora have been manually annotated into relevant and irrelevant fragments with their contained APIs.

*B. Baseline Approaches*

*(1) IR Approach*

IR approach uses an information retrieval method to find relevant fragments for APIs [3]. Cosine similarities with TF-IDF term weight are calculated between fragments and API specifications. A fragment is treated as relevant to an API, if their cosine similarity is higher than a threshold. The threshold in each tutorial is defined as follows: first, the K most similar fragments are achieved by calculating cosine similarities for each API, where K is the number of relevant fragments for this API based on the annotation. Then, the average of all K-th fragments' similarities for all the APIs is used as the threshold.

*(2) GMR Approach*

GMR approach is the seminal work to find relevant fragments for APIs [3]. Before extracting features, some text transformation operations are conducted, *e.g.*, sentence identification and part-of-speech tagging. Twenty features are defined to measure linguistic and structural characteristics between fragments and APIs. A MaxEnt classifier is trained to determine relevant fragments for APIs.

*(3) FITSEA Approach*

FITSEA approach is the state-of-the-art method to find relevant fragments for APIs [7]. It introduces some new sources to extend APIs to overcome the information mismatch between fragments and APIs. Besides, co-occurrence APIs are proposed to act as good indicators to improve the results.

*C. Evaluation Methods and Metrics*

Supervised approaches can be evaluated by the widely used Ten-Fold Cross Validation (TFCV) to measure how accurately an approach performs [34]. We employ TFCV rather than leave-one-out cross validation (LOOCV) in [3], Since the fundamental unit is the fragment-API pair in LOOCV, and a fragment containing multiple APIs is simultaneously put into both the training set and the test set, which is far away from real applications. In TFCV, we treat fragments as fundamental units to partition a tutorial into 10 subsets of equal size after identifying APIs and generating fragment-API pairs. Among the 10 subsets, one subset is used as the test set, and the other 9 subsets are used as the training set. After all the subsets are used once as the test set when cross validation is repeated 10 times, the 10 results are averaged to evaluate the performance of the approach.

In contrast, unsupervised approaches (FRAPT and IR) can determine the relevance between fragments and APIs without the overhead of training. Given a tutorial fragment or an API, unsupervised approaches can discover its relevant APIs or fragments using their designed frameworks.

In this study, we introduce *Precision*, *Recall*, and *F-Measure* to evaluate the performance of each approach. These metrics are employed by IR, GMR, and FITSEA in [3, 7] as evaluation metrics. *Precision* measures how accurate the results of experiments are. *Recall* indicates the coverage of the results. *F-Measure* balances *Precision* and *Recall*, since there is a tradeoff between them.

There are four kinds of results obtained from the experiments, namely True Positive (TP), False Negative (FN), True Negative (TN), and False Positive (FP). TPs indicate the results which correctly predict relevant fragment-API pairs as relevant. FNs indicate the results which incorrectly predict relevant fragment-API pairs as irrelevant. TNs and FPs indicate the results which predict irrelevant fragment-API pairs as irrelevant and relevant respectively. *Precision*, *Recall*, and *F-Measure* can be calculated as follows:

$$Precision = \frac{\#TP}{\#TP + \#FP} \times 100\% \quad (4)$$

$$Recall = \frac{\#TP}{\#TP + \#FN} \times 100\% \quad (5)$$

$$F\text{-}Measure = \frac{2 \times Precision \times Recall}{Precision + Recall} \times 100\% \quad (6)$$

## VI. EXPERIMENTAL RESULTS

In this section, we investigate five Research Questions (RQs) to evaluate different aspects of FRAPT.

*A. Investigation to RQ1*

**RQ1: How does the threshold $T$ influence the performance of FRAPT?**

**Motivation.** The threshold $T$ plays an important role to determine the relevance between fragments and APIs in the Relevance Identification component (see IV.F). We try to explore how the threshold $T$ influences the results and decide a good value ($T_0$) for $T$ in this RQ.

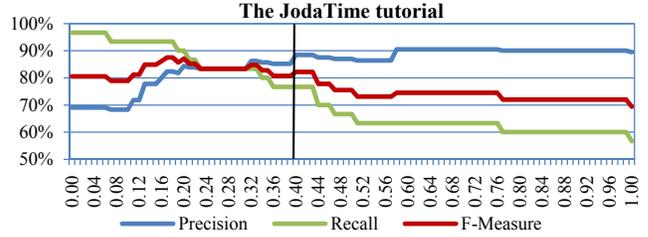
Figure 3. Results of RRAPT on JodaTime tutorial

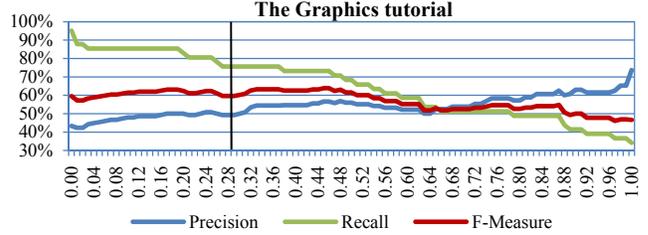
Figure 4. Results of FRAPT on Graphics tutorial

**Approach.** We select one tutorial from the two tutorial corpora respectively as cases to investigate this RQ, namely the JodaTime tutorial in McGill corpus and the Graphics tutorial in Android corpus. We adjust $T$ from 0 to 1 with a step size of 0.01 in the selected tutorials to obtain the results.

Given a tutorial, we try to find a good tutorial-specific value threshold $T_0$. By comparing the values of *Precision*, *Recall*, and *F-Measure*, we can validate whether $T_0$ is close the optimal value. $T_0$ is derived from the percentage of sentences containing APIs and their methods in the whole tutorial. This scheme is designed by the observation: Intuitively, more sentences in a fragment contain APIs and their methods, more chances the fragment is relevant to APIs. Hence, $T_0$ can be calculated as follows:

$$T_0 = 1 - \frac{\text{\# sentences in a tutorial containing APIs and their methods}}{\text{\# all the sentences in a tutorial}} \quad (7)$$

**Result.** Fig. 3 and Fig. 4 show the *Precision*, *Recall*, and *F-Measure* against $T$ for the two selected tutorials. We can see that, with the growth of $T$, the curves of *Precision* show upward trends in both tutorials. For example, when $T$ is set to 0, FRAPT achieves a *Precision* of 69.05% in the JodaTime tutorial. When $T$ comes to 1, the *Precision* rises to 89.47%. On the contrary, the curves of *Recall* show downward trends in both tutorials. Here, the *Recall* cannot achieve 100% when $T$ is set to 0, since some explanatory fragments are filtered out as non-explanatory fragments in the Fragment Filter component. For example, in the Graphics tutorial, FRAPT achieves a *Recall* of 95.12% when $T$ set to is 0, and it drops to 34.15% when $T$ is set to 1. Compared to *Precision* and *Recall*, the curves of *F-Measure* are relatively stable.

The black vertical lines in Fig. 3 and Fig. 4 show $T_0$ calculated by the formula (7) in the two tutorials, namely 0.39 and 0.29 respectively. We can see that the values of *F-Measure* almost approximate the best values using $T_0$ in the JodaTime tutorial and the Graphics tutorial. Even though FRAPT does not achieve the best *F-Measure*, the gap is trivial. For instance, FRAPT achieves the *F-Measure* of 59.62% in the Graphics tutorial, which is lower than the best value only by 4.21%.

Table V. DETAILED RESULTS OF DIFFERENT APPROACHES

| Corpus | Tutorial | Precision (%) | | | | Recall (%) | | | | F-Measure (%) | | | |
|---|---|---|---|---|---|---|---|---|---|---|---|---|---|
| | | FRAPT | FITSEA | GMR | IR | FRAPT | FITSEA | GMR | IR | FRAPT | FITSEA | GMR | IR |
| McGill Corpus | JodaTime | **85.19** | 69.00 | 58.82 | 73.00 | **76.67** | 74.17 | 50.00 | 73.00 | **80.70** | 70.24 | 54.05 | 73.00 |
| | Math Library | **84.78** | 67.89 | 52.00 | 67.00 | **73.58** | 72.70 | 49.06 | 65.00 | **78.79** | 61.53 | 50.49 | 66.00 |
| | Col. Official | 62.03 | 55.74 | **62.69** | 30.00 | 87.50 | 48.62 | 31.79 | **94.00** | 72.59 | 48.10 | 42.18 | 45.00 |
| | Col. Jenkov | 61.19 | **90.44** | 82.14 | 33.00 | **97.62** | 85.17 | 58.97 | 88.00 | 75.23 | **85.17** | 68.66 | 48.00 |
| | Smack | 77.94 | **83.38** | 70.00 | 74.00 | **94.64** | 88.33 | 93.33 | 52.00 | **85.48** | 83.90 | 80.00 | 61.00 |
| Android Corpus | Graphics | 49.21 | **50.42** | 45.60 | 35.80 | **75.61** | 42.52 | 44.50 | 67.44 | **59.62** | 43.73 | 45.04 | 46.77 |
| | Resources | 65.22 | **75.83** | 55.00 | 40.32 | **66.67** | 66.17 | 21.11 | 55.56 | 65.93 | **66.80** | 30.51 | 46.73 |
| | Data | **71.43** | 56.52 | 19.29 | 33.33 | **55.56** | 52.00 | 14.76 | 44.00 | **62.50** | 54.17 | 16.72 | 37.93 |
| | Text | 57.58 | 36.19 | **66.67** | 37.21 | **76.00** | 48.33 | 22.22 | 57.14 | **65.52** | 39.56 | 33.33 | 45.07 |

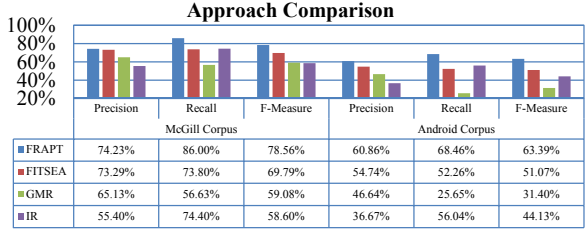

Figure 5. Average results of each approach

| | Precision | Recall | F-Measure | Precision | Recall | F-Measure |
|---|---|---|---|---|---|---|
| | McGill Corpus | | | Android Corpus | | |
| FRAPT | 74.23% | 86.00% | 78.56% | 60.86% | 68.46% | 63.39% |
| FITSEA | 73.29% | 73.80% | 69.79% | 54.74% | 52.26% | 51.07% |
| GMR | 65.13% | 56.63% | 59.08% | 46.64% | 25.65% | 31.40% |
| IR | 55.40% | 74.40% | 58.60% | 36.67% | 56.04% | 44.13% |

**Conclusion.** The threshold $T$ may influence the performance of FRAPT. The tutorial-specific threshold $T_0$ is close to the optimal value. In the following RQs, we use the threshold $T_0$ for each tutorial to conduct all the experiments.

### B. Investigation to RQ2

*RQ2: To what extent is FRAPT superior to the three baseline approaches over the two tutorial corpora?*

**Motivation.** In this RQ, we want to examine whether our unsupervised approach FRAPT can be superior to the existing approaches over the two tutorial corpora.

**Approach.** FRAPT and the three baseline approaches [3, 7] are implemented and tested over the two tutorial corpora to compare the results.

**Results.** Table V shows the detailed values of *Precision*, *Recall*, and *F-Measure* for each approach over the two tutorial corpora. We can see that different approaches behave differently over the two corpora. FRAPT shows the best *Recall* and *F-Measure* in almost all tutorials. For example, FRAPT gets the best *Precision* of 85.19% among all the approaches in the JodaTime tutorial in McGill corpus. At the same time, it also achieves the best *Recall* and *F-Measure* (76.67% and 80.70% respectively).

The average values of these evaluation metrics are presented in Fig. 5. We can see that FRAPT achieves the best results in all the conditions among all the approaches. For example, the average *Precision*, *Recall*, and *F-Measure* of FRAPT are 74.23%, 86.00%, and 78.56% respectively on McGill corpus, whereas the other approaches are all below 70% for the *F-Measure*. FRAPT improves the state-of-the-art approach FITSEA by 8.77% and 12.32% on average over the two corpora in terms of *F-Measure*. When comparing different supervised approaches, FITSEA achieves better results than GMR, *e.g.*, F-Measure of FITSEA is 69.79% over McGill corpus, whereas GMR only achieves 59.08%.

Table VI. RESULTS OF FRAGMENT FILTER

| McGill Corpus | | Android Corpus | |
|---|---|---|---|
| True Irrelevant | False Irrelevant | True Irrelevant | False Irrelevant |
| 66 | 9 | 27 | 9 |

Table VII. DETAILED RESULTS OF FRAPT AND FRAPT-FILTER

| Corpus | Tutorial | Precision (%) | | Recall (%) | | F-Measure (%) | |
|---|---|---|---|---|---|---|---|
| | | FRAPT | FRAPT-Filter | FRAPT | FRAPT-Filter | FRAPT | FRAPT-Filter |
| McGill Corpus | JodaTime | **85.19** | 60.00 | 76.67 | **80.00** | **80.70** | 68.57 |
| | Math Library | **84.78** | 65.63 | 73.58 | **79.25** | **78.79** | 71.79 |
| | Col. Official | **62.03** | 38.30 | 87.50 | **96.43** | **72.59** | 54.82 |
| | Col. Jenkov | **61.19** | 37.84 | 97.62 | **100.00** | **75.23** | 54.90 |
| | Smack | **77.94** | 75.00 | 94.64 | **96.43** | **85.48** | 84.38 |
| Android Corpus | Graphics | **49.21** | 43.24 | 75.61 | **78.05** | **59.62** | 55.65 |
| | Resources | **65.22** | 61.82 | 66.67 | **75.56** | 65.93 | **68.00** |
| | Data | **71.43** | 51.52 | 55.56 | **62.96** | **62.50** | 56.67 |
| | Text | **57.58** | 42.55 | 76.00 | **80.00** | **65.52** | 55.56 |

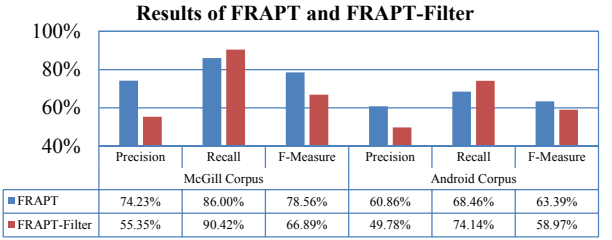

| | Precision | Recall | F-Measure | Precision | Recall | F-Measure |
|---|---|---|---|---|---|---|
| | McGill Corpus | | | Android Corpus | | |
| FRAPT | 74.23% | 86.00% | 78.56% | 60.86% | 68.46% | 63.39% |
| FRAPT-Filter | 55.35% | 90.42% | 66.89% | 49.78% | 74.14% | 58.97% |

Figure 6. Average results of FRAPT and FRAPT-Filter

**Conclusion.** As an unsupervised approach, FRAPT can achieve better results than the state-of-the-art supervised approach verified by TFCV. Considering the advantages of unsupervised approaches, it is a better choice to use FRAPT for discovering relevant tutorial fragments for APIs.

### C. Investigation to RQ3

*RQ3: What is the impact of Fragment Filter on detecting non-explanatory fragments and improving results of FRAPT?*

**Motivation.** We build a Fragment Filter to detect and filter out non-explanatory fragments (see IV.C) to address the *non-explanatory fragment identification* challenge. We try to explore its impact on FRAPT in this RQ.

**Approach.** There are two types of results, namely true non-explanatories and false non-explanatories. True non-explanatories are the results in which non-explanatory fragments are correctly marked as non-explanatory, while false non-explanatories are the results in which explanatory fragments are incorrectly marked as non-explanatory. After applying Fragment Filter, we count the number of true non-explanatories and false non-explanatories to show the effectiveness of Fragment Filter.

Table VIII. DETAILED RESULTS OF FRAPT AND FRAPT-PRONSRES

| Corpus | Tutorial | Precision (%) | | Recall (%) | | F-Measure (%) | |
|---|---|---|---|---|---|---|---|
| | | FRAPT | FRAPT-PronsRes | FRAPT | FRAPT-PronsRes | FRAPT | FRAPT-PronsRes |
| McGill Corpus | JodaTime | **85.19** | 80.00 | 76.67 | **93.33** | 80.70 | **86.15** |
| | Math Library | **84.78** | 79.31 | 73.58 | **86.79** | 78.79 | **82.88** |
| | Col. Official | **62.03** | 56.18 | 87.50 | **89.29** | **72.59** | 68.97 |
| | Col. Jenkov | **61.19** | 50.00 | 97.62 | 97.62 | **75.23** | 66.13 |
| | Smack | **77.94** | 71.62 | **94.64** | 94.64 | **85.48** | 81.54 |
| Android Corpus | Graphics | 49.21 | **53.45** | **75.61** | 75.61 | 59.62 | **62.63** |
| | Resources | **65.22** | 65.12 | **66.67** | 62.22 | **65.93** | 63.64 |
| | Date | **71.43** | 71.43 | **55.56** | 55.56 | **62.50** | 62.50 |
| | Text | **57.58** | 55.17 | **76.00** | 64.00 | **65.52** | 59.26 |

Table IX. DETAILED RESULTS OF FRAPT AND ITS VARIANTS

| Corpus | Tutorial | Precision (%) | | | Recall (%) | | | F-Measure (%) | | |
|---|---|---|---|---|---|---|---|---|---|---|
| | | FRAPT | FRAPT-Topic | FRAPT-PR | FRAPT | FRAPT-Topic | FRAPT-PR | FRAPT | FRAPT-Topic | FRAPT-PR |
| McGill Corpus | JodaTime | **85.19** | 68.29 | 67.50 | 76.67 | **93.33** | 90.00 | **80.70** | 78.87 | 77.14 |
| | Math Library | **84.78** | 82.35 | 71.64 | 73.58 | 79.25 | **90.57** | 78.79 | **80.77** | 80.00 |
| | Col. Official | **62.03** | 56.18 | 52.63 | 87.50 | **89.29** | 89.29 | **72.59** | 68.97 | 66.23 |
| | Col. Jenkov | **61.19** | 55.41 | 50.00 | 97.62 | **97.62** | 97.62 | **75.23** | 70.69 | 66.13 |
| | Smack | **77.94** | 69.23 | 68.35 | 94.64 | **96.43** | 96.43 | **85.48** | 80.60 | 80.00 |
| Android Corpus | Graphics | 49.21 | 42.35 | **43.33** | 75.61 | 87.80 | **95.12** | **59.62** | 57.14 | 59.54 |
| | Resources | **65.22** | 60.38 | 49.25 | 66.67 | 71.11 | **73.33** | **65.93** | 65.31 | 58.93 |
| | Data | **71.43** | 48.89 | 51.06 | 55.56 | 81.48 | **88.89** | 62.50 | 61.11 | **64.86** |
| | Text | **57.58** | 55.88 | 50.00 | 76.00 | 76.00 | **88.00** | **65.52** | 64.41 | 63.77 |

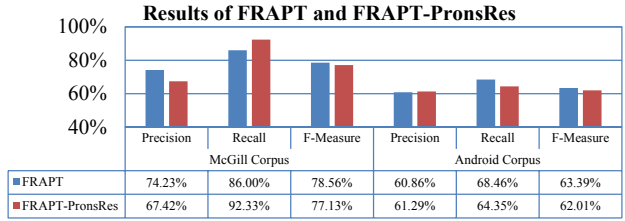

Figure 7. Average results of FRAPT and FRAPT-PronsRes

| | Precision | Recall | F-Measure | Precision | Recall | F-Measure |
|---|---|---|---|---|---|---|
| | McGill Corpus | | | Android Corpus | | |
| FRAPT | 74.23% | 86.00% | 78.56% | 60.86% | 68.46% | 63.39% |
| FRAPT-PronsRes | 67.42% | 92.33% | 77.13% | 61.29% | 64.35% | 62.01% |

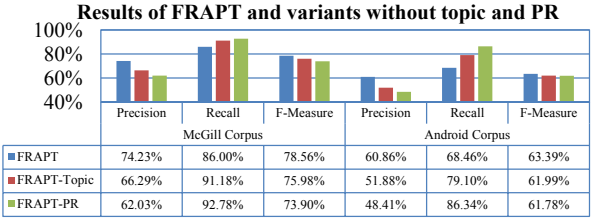

Figure 8. Average results of FRAPT and its variants

| | Precision | Recall | F-Measure | Precision | Recall | F-Measure |
|---|---|---|---|---|---|---|
| | McGill Corpus | | | Android Corpus | | |
| FRAPT | 74.23% | 86.00% | 78.56% | 60.86% | 68.46% | 63.39% |
| FRAPT-Topic | 66.29% | 91.18% | 75.98% | 51.88% | 79.10% | 61.99% |
| FRAPT-PR | 62.03% | 92.78% | 73.90% | 48.41% | 86.34% | 61.78% |

We also compare FRAPT against a variant of FRAPT, namely FRAPT-Filter which removes Fragment Filter and keeps the other components the same. By comparing FRAPT against FRAPT-Filter, the impact of Fragment Filter can be further shown.

**Results.** As shown in Table VI, Fragment Filter correctly marks 66 and 27 true non-explanatories for McGill corpus and Android corpus and incorrectly marks 9 false non-explanatories respectively. Fragment Filter can detect a great number of non-explanatory fragments accurately and only falsely filter out a fraction of explanatory fragments.

Table VII and Fig. 6 show the detailed and average results of FRAPT against FRAPT-Filter. We can see from Table VII that Fragment Filter can largely improve *Precision* and *F-Measure*. For instance, FRAPT obtains *F-Measure* of 80.70% in the JodaTime tutorial, whereas FRAPT-Filter only achieves 68.57%. As shown in Fig. 6, introducing Fragment Filter will effectively improve *Precision* and *F-Measure*. For example, FRAPT achieves an average *F-Measure* of 78.56% on McGill corpus, whereas FRAPT-Filter only achieves 66.89% in the same situation. We can also find similar phenomenon on Android corpus. FRAPT-Filter can achieve better *Recall* on average, whereas it shows far worse *Precision* and *F-Measure*. Overall, FRAPT balances *Precision* and *Recall* and shows better results.

**Conclusion.** Fragment Filter can detect a great number of non-explanatory fragments at the cost of a small fraction of explanatory fragments. It can balance *Precision* and *Recall* effectively, thus improve the results of FRAPT.

*D. Investigation to RQ4*

*RQ4: How does pronoun and variable resolution impact on the performance of FRAPT?*

**Motivation.** We perform pronoun and variable resolution for each fragment in Fragment Parser to address the *pronoun and variable resolution* challenge. To test whether resolution is effective, we set up this RQ.

**Approach.** We define a variant of FRAPT, namely FRAPT-PronsRes which deletes the subcomponent of pronoun and variable resolution from FRAPT. We compare the results of FRAPT against FRAPT-PronsRes to show the impact of pronoun and variable resolution in this RQ.

**Results.** Table VIII and Fig. 7 show the detailed results and average results of FRAPT and FRAPT-PronsRes respectively. We can see from Table VIII that applying pronoun and variable resolution can improve the results. For example, FRAPT improves *F-Measure* by 6.26% in the Text tutorial of Android corpus. We can see from Fig. 7 that the average *F-Measures* of FRAPT-PronsRes over the two tutorial corpora are 77.13% and 62.01% and FRAPT improves it by 1.43% and 1.38% respectively.

We find that using pronoun and variable resolution achieves different improvements in distinct tutorial corpora. There are two potential reasons leading to this phenomenon. First, different tutorials may contain different numbers of pronouns and variables, hence different improvements are achieved. Second, pronoun and variable resolution will enhance the influence of relevant APIs. At the same time, the influence of irrelevant APIs may also be strengthened.

**Conclusion.** Pronoun and variable resolution is an effective text transformation operation to improve the results.

*E. Investigation to RQ5*

*RQ5: Does FRAPT achieve better results by aggregating topic model and PageRank algorithm than by any of them alone?*

**Motivation.** FRAPT finds relevant fragments for APIs by aggregating both topic model and PageRank algorithm. To explore whether leveraging both of them can improve the results of FRAPT than any of them alone, we set up this RQ.

**Approach.** In the same way as the previous RQs, we set up two variants of FRAPT. The first one is FRAPT-Topic which only considers the correlation scores from PageRank algorithm. The second variant is FRAPT-PR which only takes the correlation scores from topic model into account.

**Results.** Table IX shows the values of *Precision*, *Recall*, and *F-Measure* over the two tutorial corpora. We can see from the table that, in almost all the situations, FRAPT achieves the best results compared to its variants in terms of *Precision* and *F-Measure*. Taking the Smack tutorial in McGill corpus as an example, the *Precision* is 77.94% in FRAPT, whereas FRAPT-Topic and FRAPT-PR achieve 69.23% and 68.35% respectively. When considering *F-Measure*, FRAPT outperforms FRAPT-Topic and FRAPT-PR by 4.88% and 5.48% respectively. When taking *Recall* into account, even though FRAPT shows a slight decline, it can better balance *Precision* and *Recall* to achieve significant improvement. From Fig. 8 which shows the average values, we can see that FRAPT shows its advantage.

The reason why aggregating topic model and PageRank algorithm could improve the results may be that, topic model can find semantic relationships between fragments and APIs, while PageRank algorithm tries to find lexical and structural connections between them. By aggregating both topic model and PageRank algorithm, FRAPT could better capture the relevance between fragments and APIs.

**Conclusion.** By aggregating both topic model and PageRank algorithm, FRAPT can better detect relevant fragments for APIs.

## VII. THREATS TO VALIDITY

### A. Threats to Internal Validity

Internal validity is the degree of our approach to link independent variables to dependent variables. In the experiments, we have already investigated the influences of several important components, including Fragment Filter and pronoun and variable resolution, etc. From the results of the RQs, we can find that they are effective components. Besides, topic model and PageRank algorithm have been applied to many areas, and their stability and scalability have been proved. As a result, this threat has been taken into consideration and minimized as many as possible in FRAPT.

### B. Threats to External Validity

External validity is the ability to which the conclusion of the experiments can be generalized to other corpora and research settings. In this study, we introduce all the tutorial corpora we can obtain. These tutorials are related to Java and Android APIs, and cover many different topics with different sizes and origins [35]. As a result, they are representative tutorials for research study. It is unclear how our approach performs when applying to other tutorials. In the future, we plan to introduce more tutorials to verify FRAPT.

## VIII. RELATED WORK

There are a lot of API documentations helping developers to use APIs, *e.g.*, API specifications, API tutorials, forums, and blogs [17-19, 25]. A series of works exist in the literature related to API documentations [26-33]. We mainly introduce three research topics, *i.e.* content classification, information enhancement, and error detection.

Content classification aims to analyze the content of API documentations to improve the efficiency of consulting API documentations. Maalej and Robillard [20] developed a taxonomy of knowledge types to study the nature of knowledge in API reference documentations. Monperrus et al. [21] conducted an empirical study on the directives, and 23 kinds of directives were developed and discussed. Dekel et al. [22] developed and released an Eclipse plugin named *eMoose* to show the associated directives for methods.

Information enhancement augments API documentations by some knowledge units. Treude and Robillard [23] proposed SISE to augment API documentations with insight sentences from Stack Overflow. Kim et al. [24] proposed an approach to enrich API documentations with code examples. The results of user study showed that the enriched API documentations can improve the productivity of developers.

Error detection attempts to find errors in API documentations. Zhong and Su [19] formulated a set of inconsistencies and combined natural language processing as well as code analysis techniques to find these inconsistencies, and more than 1000 errors were found and reported.

As typical API documentations, we focus on API tutorials in this study. Different from the above research topics, we attempt to break up API tutorials into fragments and find relevant fragments for APIs.

## IX. CONCLUSION AND FUTURE WORK

It remains a challenging issue to discover relevant fragments for APIs. In this study, we propose an unsupervised approach, namely FRAPT. Combined with two fragment examples, we demonstrate how each component works in FRAPT. We compare FRAPT against three existing approaches over two publicly open tutorial corpora. The results show that FRAPT achieves better results than the state-of-the-art approach. Besides, some key components in FRATP are also evaluated to show their effectiveness.

For future works, we try to improve FRAPT in several directions. First, we plan to address the threats to external validity by introducing more tutorial corpora to validate FRAPT. Second, we try to adapt FRAPT to discover relevant fragments for different levels of APIs, *e.g.*, method level. Third, we try to develop an Eclipse plugin encapsulating FRAPT to facilitate learning and using APIs for developers.

ACKNOWLEDGMENT

We thank the reviewers for their comments. This work is supported by the New Century Excellent Talents in University under Grant NCET-13-0073, the National Natural Science Foundation of China under Grants 61370144, 61403057, and 61602258, the Fundamental Research Funds for the Central Universities under Grant DUT14YQ203.

REFERENCES

[1] M. Morisio, M. Ezran, and C. Tully, "Success and Failure Factors in Software Reuse," IEEE Transactions on Software Engineering, vol. 28, pp. 340-357, April 2002.

[2] B. Dagenais and M. P. Robillard, "Recovering Traceability Links between an API and its Learning Resources," In Proceedings of the 34th International Conference on Software Engineering (ICSE 12), 2012, pp. 47-57.

[3] G. Petrosyan, M. P. Robillard, and R. de. Mori, "Discovering Information Explaining API Types using Text Classification," In Proceedings of the 37th International Conference on Software Engineering (ICSE 15), 2015, pp. 869-879.

[4] M. P. Robillard, "What Makes APIs Hard to Learn? Answers from Developers," IEEE software, vol. 26, pp. 27-34, November 2009.

[5] M. P. Robillard and R. Deline, "A Field Study of API Learning Obstacles," Empirical Software Engineering, vol. 16, pp. 703-732, December 2011.

[6] L. Ponzanelli, G. Bavota, A. Mocci, et al. "Too Long; Didn't Watch!: Extracting Relevant Fragments from Software Development Video Tutorials," In Proceedings of the 38th International Conference on Software Engineering (ICSE 16). 2016, pp. 261-272.

[7] H. Jiang, J. X. Zhang, X. C. Li, et al. "A More Accurate Model for Finding Tutorial Segments Explaining APIs," In Proceedings of the 23rd IEEE International Conference on Software Analysis, Evolution, and Reengineering (SANER 16), 2016, pp. 157-167.

[8] E. Alpaydin, Introduction to Machine Learning (Adaptive Computation and Machine Learning). The MIT Press, 2004.

[9] R. O. Duda, P. E. Hart, and D. G. Stork, Unsupervised Learning and Clustering. Pattern classification (2nd). Wiley, 2001.

[10] V. Stoyanov, C. Cardie, N. Gilbert, E. Riloff, D. Buttler, and D. Hysom, "Coreference Resolution with Reconcile," In Proceedings of the Conference of the 48th Annual Meeting of the Association for Computational Linguistics (ACL 10), 2010, Short Paper.

[11] Alias-i. 2008. LingPipe 4.1.0. http://alias-i.com/lingpipe.

[12] A. Panichella, B. Dit, R. Oliveto, M. D. Penta, D. Poshynanyk and A. D. Lucia, "How to Effectively Use Topic Models for Software Engineering Tasks? An Approach Based on Genetic Algorithms," In Proceedings of the 35th International Conference on Software Engineering (ICSE 13). 2013, pp. 522-531.

[13] S. W. Thomas, Mining Software Repositories with Topic Models. School of Computing. Queen's University, Canada. 2012.

[14] Stanford Topic Modeling Toolbox. http://nlp.stanford.edu/software/tmt/tmt-0.2/.

[15] A. Alon and M. Tennenholtz, "Ranking Systems: The PageRank Axioms," In Proceedings of the 6th ACM conference on Electronic commerce, 2005, pp. 1-8.

[16] A. N. Langville and C. D. Meyer, Google's PageRank and Beyond: The Science of Search Engine Rankings. Princeton University Press. ISBN 0-691-12202-4. 2006.

[17] L. Shi, H. Zhong, T. Xie, et al. "An Empirical Study on Evolution of API Documentation," In Proceedings of the International Conference on Fundamental Approaches to Software Engineering (FASE 11), 2011, pp. 416-431.

[18] B. Dagenais and M. P. Robillard, "Using Traceability Links to Recommend Adaptive Changes for Documentation Evolution," IEEE Transactions on Software Engineering, vol. 40, pp. 1126-1146, November 2014.

[19] H. Zhong and Z. D. Su, "Detecting API Documentation Errors," In Proceedings of the 2013 ACM SIGPLAN international conference on Object oriented programming systems languages & applications (OOPSLA 13), 2013, pp. 803-816.

[20] W. Maalej and M. P. Robillard, "Patterns of Knowledge in API Reference Documentation," IEEE Transactions on Software Engineering, vol. 39, pp. 1264-1283, September 2013.

[21] M. Monperrus, M. Eichberg, E. Tekes, et al. "What Should Developers Be Aware Of? An Empirical Study on the Directives of API Documentation," Empirical Software Engineering, vol. 17, pp. 703-737, December 2012.

[22] U. Dekel and J. D. Herbsleb, "Improving API Documentation Usability with Knowledge Pushing," In Proceedings of the 31st International Conference on Software Engineering (ICSE 09), 2009, pp. 320-330.

[23] C. Treude and M. P. Robillard, "Augmenting API Documentation with Insights from Stack Overflow," In Proceedings of the 38th International Conference on Software Engineering (ICSE 16), 2016, pp. 392-403.

[24] J. Kim, S. Lee, S. W. Hwang, et al. "Enriching Documents with Examples: A Corpus Mining Approach," ACM Transactions on Information Systems (TOIS), vol. 31, January 2013.

[25] S. Subramanian, L. Inozemtseva, and R. Holmes, "Live API Documentation," In Proceedings of the 36th International Conference on Software Engineering (ICSE 14), 2014, pp. 643-652.

[26] H. Zhong, L. Zhang, T. Xie, and H. Mei, "Inferring Resource Specifications from Natural Language API Documentation," In proceedings of The 2009 International Conference on Automated Software Engineering (ASE 09), 2009, pp. 307-318.

[27] D. Kramer, "API Documentation from Source Code Comments: A Case Study of Javadoc," In Proceedings of the 17th Annual International Conference on Computer Documentation (SIGDOC 99), 1999, pp. 147-153.

[28] M. P. Robillard and Y. B. Chhetri, "Recommending Reference API Documentation," Journal of Empirical Software Engineering, vol. 20, pp. 1558-1586, December 2015.

[29] D. Hoffman and P. Strooper, "API Documentation with Executable Examples", Journal of Systems and Software, vol. 66, pp. 143-156, May 2003.

[30] Y. Wu, L. W. Mar, and H. C. Jiau, "CoDocent: Support API Usage with Code Example and API Documentation," In Proceedings of 5th International Conference on Software Engineering Advances (ICSEA 10), 2010, pp. 135-140.

[31] E. D. Ekoko and M. P. Robillard, "Asking and Answering Questions about Unfamiliar APIs: An Exploratory Study," In Proceedings of the 34th International Conference on Software Engineering (ICSE 12), 2012, pp. 266-276.

[32] C. Chen and K. Zhang, "Who Asked What: Integrating Crowdsourced FAQs into API Documentation," In Companion Proceedings of the 36th International Conference on Software Engineering (ICSE 14), pp. 456-459.

[33] G. Uddin, B. Dagenais, and M. P. Robillard, "Analyzing Temporal API Usage Patterns," In Proceedings of International Conference on Automated Software Engineering (ASE 11), 2011, pp. 456-459.

[34] P. C. Rigby and M. P. Robillard, "Discovering Essential Code Elements in Informal Documentation," In Proceedings of the 35th International Conference on Software Engineering (ICSE 13), 2013, pp. 832-841.

[35] L. Mastrangelo, L. Ponzanelli, A. Mocci, M. Lanza, M. Hauswirth, and N. Nystrom, "Use at Your Own Risk: The Java Unsafe API in the Wild," In Proceedings of the 30th ACM SIGPLAN Conference on Object-Oriented Programming, Systems, Languages, and Applications (OOPSLA 15), 2015, pp. 695-710.

[36] Powers and M. W. David, "The Problem with Kappa," In Proceedings of Conference of the European Chapter of the Association for Computational Linguistics (EACL 12), 2012.

[37] http://docs.oracle.com/javase/tutorial/essential/index.html.

[38] http://cs.mcgill.ca/~swevo/icse2015/.

[39] http://oscar-lab.org/paper/API/.

[40] http://joda-time.sourceforge.net/userguide.html.

[41] http://developer.android.com/guide/topics/graphics/index.html.